\documentclass[10pt,twocolumn]{article} 
\usepackage{simpleConference}
\usepackage{times}
\usepackage{graphicx}
\usepackage{amssymb}
\usepackage{url,hyperref}

\usepackage{cite}
\usepackage{hyperref}
\usepackage{amsmath,amssymb,amsfonts}
\usepackage{algorithmic}
\usepackage{graphicx}
\usepackage{xcolor}
\usepackage{xspace}
\usepackage{listings}
\usepackage{balance}
\usepackage{todonotes}

\definecolor{codegreen}{rgb}{0,0.6,0}
\definecolor{codegray}{rgb}{0.5,0.5,0.5}
\definecolor{codepurple}{rgb}{0.58,0,0.82}
\definecolor{backcolour}{rgb}{0.95,0.95,0.92}

\lstdefinestyle{mystyle}{
    backgroundcolor=\color{backcolour},   
    commentstyle=\color{codegreen},
    keywordstyle=\color{magenta},
    numberstyle=\tiny\color{codegray},
    stringstyle=\color{codepurple},
    basicstyle=\ttfamily\footnotesize,
    breakatwhitespace=false,         
    breaklines=true,                 
    captionpos=b,                    
    keepspaces=true,                 
    numbersep=5pt,                  
    showspaces=false,                
    showstringspaces=false,
    showtabs=false,                  
    tabsize=2
}

\usepackage{csquotes}
\usepackage{cleveref}

\usepackage{dblfloatfix}

\usepackage[numbers]{natbib}
\usepackage[most]{tcolorbox}

\newcommand{\msrflakiness}{\textsf{\text{msr4flakiness}}\xspace}

\def\BibTeX{{\rm B\kern-.05em{\sc i\kern-.025em b}\kern-.08em
    T\kern-.1667em\lower.7ex\hbox{E}\kern-.125emX}}
    
\begin{document}

\title{What is the Vocabulary of Flaky Tests? \\An Extended Replication}

\author{
    \textbf{Bruno Henrique Pachulski Camara} \\
    \textit{Department of Computer Science} \\
    \textit{Federal University of Paraná} \\
    Curitiba, PR, Brazil \\
    \texttt{bhpachulski@ufpr.br} \\

\and

    \textbf{Marco Aurélio Graciotto Silva}\\
    \textit{Department of Computing} \\
    \textit{Federal University of Technology - Paraná}\\
    Campo Mourão, PR, Brazil \\
    \texttt{magsilva@utfpr.edu.br} \\
    
\and

    \textbf{Andre Takeshi Endo} \\
    \textit{Department of Computing} \\
    \textit{Federal University of Technology - Paraná} \\
    Cornélio Procópio, PR, Brazil \\
    \texttt{andreendo@utfpr.edu.br} \\

\and

    \textbf{Silvia Regina Vergilio} \\
    \textit{Department of Computer Science} \\
    \textit{Federal University of Paraná} \\
    Curitiba, PR, Brazil \\
    \texttt{silvia@inf.ufpr.br} \\

}

\maketitle

\begin{abstract}
Software systems have been continuously evolved and delivered with high quality due to the widespread adoption of automated tests. 
A recurring issue hurting this scenario is the presence of flaky tests, a test case that may pass or fail non-deterministically. 
A promising, but yet lacking more empirical evidence, approach is to collect static data of automated tests and use them to predict their flakiness. 
In this paper, we conducted an empirical study to assess the use of code identifiers to predict test flakiness. To do so, we first replicate most parts of the previous study of Pinto~et~al.~(MSR~2020). This replication was extended by using a different ML Python platform (Scikit-learn) and adding different learning algorithms in the analyses. Then, we validated the performance of trained models using  datasets with other flaky tests and from different projects. 
We successfully replicated the results of Pinto~et~al.~(2020), with minor differences using Scikit-learn; different algorithms had performance similar to the ones used previously. Concerning the validation, we noticed that the recall of the trained models was smaller, and classifiers presented a varying range of decreases. This was observed in both intra-project and inter-projects test flakiness prediction. 

\textbf{Kewords:} test flakiness, regression testing, replication studies, machine learning
\end{abstract}

\section{Introduction}

In regression testing, automated tests are run to validate whether changes and/or bug fixes do not have a negative impact on the software.  Nevertheless, not all test failures in regression imply on faults in the production code~\cite{HerzigNagappan2015}.
Some tests have an intermittent behavior: they may pass or fail when executed in the same software version.
Those tests cannot be trusted and are known as flaky tests~\cite{Luo-etal:2014}.

Unfortunately, flaky tests are common.  
\citet{Eck-etal:2019} surveyed 121 practitioners from Mozilla Foundation and reported test flakiness as a moderate to critical issue. Moreover, they found 58\% of professionals had faced this problem in the last month. In such cases, developers may
spend important resources in analyzing failures that are due to flaky
tests and not to actual problems in production code, with concrete
impact on productivity and costs. Practitioners got now used to
rerun each newly observed failure several times, to ascertain that it
was a genuine regression failure and not an intermittent one~\cite{Pinto-etal:2020}.

This problem has brought attention form practitioners~\cite{Micco:2016, Fowler:2011, Palmer:2019, Goddard:2018} and researchers~\cite{Zolfaghari-etal:2020}. We can find some approaches to detect flaky tests. Most of them require that a lot of test cases are executed many times, ~\cite{Bell-etal:2018, Lam-etal:2019}.  Because of this, static approaches have been proposed. They usually employ  Machine Learning (ML) techniques~\cite{HerzigNagappan2015, TariqKing-etal:2018, Pinto-etal:2020} and are less costly. Among such approaches, we would like to highlight the one proposed by Pinto et al.~\cite{Pinto-etal:2020} published in the Mining Software Repositories Conference (MSR2020).  The approach automatically identifies flakiness based on a more comprehensive set of predictors than the others, and this is the reason why this study was chosen by us to be replicated.  In addition to common characteristics of the test cases such as number of lines of code and occurrence of certain Java keywords, the used set of predictors is built based on the conjecture that there are some patterns in the test case code that could be used to automatically recognize flaky tests.  Then a set of tokens are extracted and post-processed by applying some  Natural Language Processing (NLP) techniques to compile a vocabulary of flaky tests.

Moreover, the work of Pinto et al. constructed and made available a dataset of flaky and non-flaky tests by running every test case, in a set of 64k tests, 100 times (6.4 million test executions).  To this dataset, five ML algorithms implemented by the Weka framework~\cite{Witten99weka:practical} were applied to predict flakiness with  an F-measure of 0.95. The best prediction performance was obtained when using Random Forest and Support Vector Machine (SVM). In terms of the code identifiers that are most strongly associated with test flakiness, the authors found that words \textit{job}, \textit{action}, and \textit{services} are commonly associated with flaky tests.

The results obtained by the original study are very promising, but we identified some threats regarding algorithms used and the generalization of the results to other projects.  Then some questions not answered by the original work  can be posed: 1) ``Can we obtain similar results by using other different algorithms or the same algorithms with distinct implementations?'', and  2) ``Are the results valid for other datasets including different projects?''. The goal of our replication study is to address such threats and questions.  

Replication Studies have been gaining importance in software engineering. Works and reviews on this subject show an increasing number of replications published ~\cite{Bezerra15,Kitchenham08,SHEPPERD2018120,Silva12}. Many conferences have included tracks dedicated to this subject. The community has highlighted the importance of producing adequate documentation to allow replication. Shull et al.~\cite{Shull-etal:2008} identified the types of replications. When researchers apply the same procedures to answer the same research questions as closely as possible the replication is called exact.  When researchers investigate the same research question by using a different experimental procedure, it is called conceptual. Internal replications are conducted by the original researchers and  external ones by an independent group~\cite{Silva12}. They are dependent replications when researchers attempt to keep all the conditions of the experiment the same or very similar. Otherwise they are independent replications, when researchers deliberately vary one or more major aspects of the conditions of the experiment. 

In this sense, our replication is external and exact. To answer our questions we applied the same procedures of original study, with some variations in the experimental conditions: research questions, algorithms, and datasets. First, we conducted a dependent replication, by adopting the same datasets and algorithms from framework Weka. This replication ensured a correct understanding of the original setup and   confirmed  original results. After, we performed an independent replication composed of two parts. In the first part, we used the same dataset but varying the algorithms implementation  by adopting the framework  Scikit-learn~\cite{Pedregosa-etal:2011} and three additional algorithms. In the second part, we extended original work with additional research questions and performed cross-project validations to assess the performance of the trained models with different datasets, for intra- and inter-projects test flakiness prediction. This allows evaluating the generalization of the original results in real scenarios.

The main contributions of this paper are the following:
\begin{itemize}
    \item 
    Evaluation of three new algorithms: the results obtained by our replication using other ML framework such as Scikit-learn confirms through some evidence that the approach of original work, based on static detection of flaky tests, is effective. One of the new algorithms added, Logistic Regression, obtained the best value of recall, and best performance in the cross-project validation.
    \item Campbell and Stanley~\cite{Campbell-Stanley:1963} argue that experiments need to be replicated in different contexts, at different times, and under different conditions before they can produce generalizable knowledge. Thus, replications like ours help investigating if the vocabulary of flaky tests obtained by the original work remains valid and can directly be transferred for other contexts like intra- and inter project test flakiness prediction.
    \item  Implications of our findings to help researchers and developers in the challenging task of identifying flaky tests. Such implications may raise future research directions.
    \item A new repository  with the procedures, datasets, and scripts generated from this replication, made available at \url{https://github.com/bhpachulski/ICPC-RENE-Paper}.

\end{itemize}

The paper is organized as follows. Section~\ref{sec:relatedWork} reviews related work on flaky tests. Section~\ref{sec:original} describes details of the original study following replication guidelines~\cite{Carver10}. Section~\ref{sec:setup} contains the setup of our replication study. Section~\ref{sec:results} presents and analyses the obtained results. Section~\ref{sec:threats} presents the main threats of our study. Section~\ref{sec:discussion} discusses some implications of our results. Section~\ref{sec:conclusions} presents our final remarks and concludes the paper.

\section{Related Work}
\label{sec:relatedWork}

Test flakiness has a negative impact on the software development process: on one hand, debugging a yet-unknown flaky test may demand a huge effort as it is not actually a bug in the software; on the other hand, prematurely labeling (incorrectly) a test as flaky would let a bug escape to production and may harm the end-users. 

Flaky tests are not only caused by changes in the software~\cite{Luo-etal:2014, Eck-etal:2019}. 
For instance, \citet{Lam-etal:2019} observed that more than half of analyzed flaky tests are order-dependent (OD), though other causes have been investigated~\cite{Luo-etal:2014}: asynchronous executions (Async Wait), concurrent execution (Concurrency), leakage resource (Resource Leak), communication (Network), tests based on date and time (Time), reading and writing files (IO), use of random data (Randomness), operations with floating-point numbers (Floating Point Operations) and use of unordered collections (Unordered Collections). 

This problem has brought attention from practitioner and researchers. We noticed efforts on this subject in the industry~\cite{Micco:2016, Fowler:2011, Palmer:2019, Goddard:2018}. In the literature, we find works dedicated to this subject as reported by a recent review~\cite{Zolfaghari-etal:2020}. Dynamic approaches to detect flaky tests require re-execution of test cases usually fixing the number of times they will be executed~\cite{Lam-etal:2019, Shi-etal:2019}. This is expensive and error-prone; the determination of this number is challenging, and an inadequate choice can lead to false negatives. As alternative, static approaches have been proposed~\cite{Luo-etal:2014, Eck-etal:2019, Shi-etal:2019, Pinto-etal:2020, Atif-etal:2017, TariqKing-etal:2018}. Most of these approaches are based on Machine Learning~\cite{Atif-etal:2017,TariqKing-etal:2018,Pinto-etal:2020}. These last ones, more related to our replication study, are described below.

\citet{Atif-etal:2017} propose to reduce the workload of the Google \textit{Test Automation Platform} (TAP), avoiding the execution of tests with low probability of failure. Another goal is to present to developers insights about the project they are developing. 
As such, developers may take preemptive measures to prevent breakages.
The proposed approach is based on a feature vector composed by Continuous Integration tool (CI), transitions from PASSED-to-FAILED, fixes (FAILED-to-PASSED), and programmers' activity in the version control tool. 
In this study, 2.07\% of tests passed or failed at least once: 1.23\% of them revealed faults introduced by developers, and 0.84\%, namely 46,694 tests were flaky.
The approach made it possible to reduce the number of executed test cases, without neglecting the fault detection capabilities. 

\citet{TariqKing-etal:2018} introduce an ML-based approach based on Bayesian networks to predict flaky tests. The problem is modeled like a disease in which symptoms can identify a flaky test. 
The authors propose a map of symptoms and causes, using static and dynamic metrics like Complexity Metrics (Test Assertion Count, Test Class/Method Size, Depth of Inheritance Tree), Implementation Coupling Metrics (Coupling Between Objects and Selector Stability Index), Non-Determinism Metrics (Cyclomatic Complexity and Explicit Wait Count), Performance Metrics (Average Execution Time), and General Stability Metrics (Failure Rate and Flip Rate). 
The evaluation was conducted with UI tests of a proprietary Web system; the tests are executed in a CI environment and five teams took part in the study. 

After three months, the supporting tool helped to reduce the test flakiness, in some cases up to 60\%. 
Overall, the accuracy of approach's prediction was 65.7\%. The features (symptoms) with the best prediction capabilities were High Test Complexity (88\%) and Test Size (82\%).

The approaches mentioned above present some advantages. The model derived is capable to predict flakiness with less cost, without re-executions of test cases. But among these approaches, the one from Pinto et al.~\cite{Pinto-etal:2020} includes a more comprehensive set of predictors and presented promising results. Such an approach is the focus of our replication study and is detailed in the next section.

\section{Original Study}
\label{sec:original}

The main hypothesis of the original study was that there are some syntactical patterns in the code of flaky test that can be used by NLP techniques to predict flakiness without code executions. Such patterns constitute the vocabulary of flaky tests.  Below, we describe the main items regarding this study, following some guidelines for replication studies proposed by  Carver et al.~\cite{Carver10}.

\subsection{Approach}

\begin{figure}[htbp]
    \centering
    \includegraphics[width=\columnwidth]{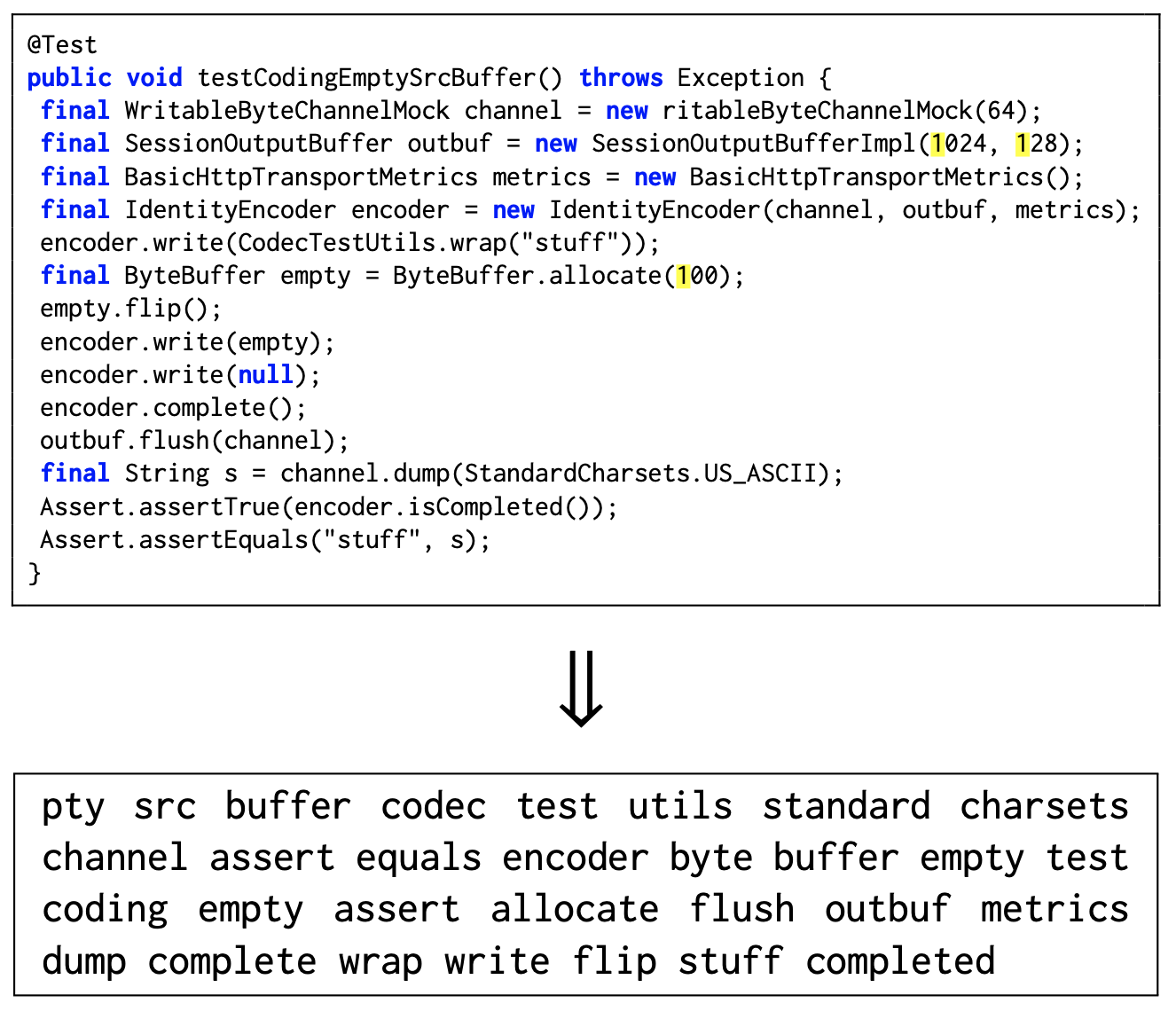}
    \caption{An example of test case and its tokenized result (extracted from~\citet{Pinto-etal:2020}.)}
    \label{fig:tokens}
\end{figure}

The adopted approach can be performed statically and includes the following  steps: 
\begin{enumerate}
    \item  Extraction of tokens:  identifiers are extracted from test code labeled as flaky or non-flaky; 

    \item  Processing of tokens: NLP techniques are applied, such as identifier splitting, stemming (removal of the suffix from a word), and stop word removal, to turn the extracted identifiers into tokens to be used as input for text classification algorithms. The identifiers are split using their camel-case syntax, and all resulting tokens are converted to lower case. \Cref{fig:tokens} contains an example of test case extracted from the original study~\cite{Pinto-etal:2020} and its corresponding set of extracted tokens.  The information about the tokens is then augmented with three numerical features, acting as proxies of code complexity:  LOC: number of lines of code of the test case; keywords: the number of occurrences of each one of the 56 Java keywords the test case contains; and keyword count: total number of Java keywords present in the test case; 

    \item  Flakiness prediction: five classifiers are applied on the resulting dataset using Weka~\cite{Witten99weka:practical}: Random Forest, Decision Tree (DT), Naive Bayes, Support Vector Machine (SVM), and Nearest Neighbour. 
\end{enumerate}

 \subsection {Research questions}
 
 The original study investigated four research questions (RQs). 
 
 \begin{itemize}
    \item \textbf{RQ1.} How prevalent and elusive are flaky tests? The goal of this question was twofold: 1) to confirm that flaky tests are very common in regression test suites, and 2) to analyse if there is an ideal number of reruns to be used to find flakiness. 
    
    \item \textbf{RQ2.}  How accurately can we predict test flakiness based on source code identifiers in the test cases? The goal of this question was to evaluate the performance of classifiers to predict test  based on the source code identifiers without re-execution of the test suites.
   
    \item \textbf{RQ3.}  What value do different features add to the classifier?  The goal of this question was to evaluate the impact of the processing step (i.e. stemming and stop word removal) in the classifier performance, as well as some choices made by the authors, such as converting identifiers to numeric features and splitting identifiers.
    
    \item \textbf{RQ4.}  Which test code identifiers are most strongly associated
with test flakiness? The goal was to identify the test code identifiers which are more related to flakiness in order to help development, code review, and debugging tasks.
 \end{itemize}

\subsection{Dataset}

The dataset was built based on 24 DeFlaker projects~\cite{Bell-etal:2018}. As this set of projects contains only information about flaky test cases, the authors decided to execute the test suites of each project 100 times. The test case was flagged as non-flaky if a consistent outcome  was obtained across all executions.  In the end, an imbalanced set was obtained with a greater number of non-flaky tests. To deal with this problem, a number of non-flaky tests was selected, equal to the number of flaky tests in the original DeFlaker data set. This selection was performed in a way that the median sizes (in number of lines of code) of flaky and non-flaky tests were nearly the same.  The data produced as result is available at: \url{https://github.com/damorimRG/msr4flakiness/}. This lab package, since now  called \msrflakiness, is one of the datasets used in our replication study.

\subsection{Evaluated metrics}

To evaluate the performance of the classifiers, the authors split the data set into 80\% for training and 20\% for testing.  They used  standard metrics of precision (the number of correctly classified flaky tests divided by the total number of tests that are classified as flaky), recall (the number of correctly classified flaky tests divided by the total number of actual flaky tests in the test set), and F$_1$-Score (the harmonic mean of precision and recall),  MCC (Matthews correlation coefficient) and AUC (area under the ROC curve). MCC measures the correlation between predicted classes (i.e., flaky vs. non-flaky) and ground truth, and AUC measures the area under the curve which visualizes the trade-off between true positive rate and false positive rate.

\subsection{Summary of the results}
As a result of RQ1, the authors found a low number of flaky tests, but flakiness is relatively common in IO-related projects. They concluded that  detecting
flakiness with test reruns is challenging. The number of executions may impact the detection since around 70\% of the test cases identified as flaky passed in more than 90\% of the runs.

Answering RQ2, all classifiers achieved very good performance in distinguishing
flaky test cases from non-flaky test cases.  Random Forest
achieved the best precision (0.99), the SVM
classifier slightly outperformed Random Forest in terms of recall
(0.92). Overall, in terms of F$_1$-Score, Random Forest achieved the
best performance, but all classifiers achieved an F$_1$-Score of at least
0.85. The results are presented in \Cref{sec:results} for comparison. 

RQ3 investigates the performance impact of the different features used in the classifiers. To answer this question, the authors used the best classifiers: Random Forest (best precision and F$_1$-Score) and Support Vector Machine (best recall). The features evaluated were:  no stemming, no stop word removal, no lower casing,  no identifier splitting, only split identifier, no LOC, no Java keywords, no identifiers.
The evaluation showed the impact on the classifier performance of most pre-processing steps was negligible. The only large impact was observed for Random Forest when 
Java keywords were included as tokens, but not identifier names. In this case, the
performance would drop from an F$_1$-Score of 0.95 to 0.79. For SVM, not splitting identifiers reduced the F$_1$-Score from 0.93 to 0.89 and not considering identifiers
at all reduced it to 0.74.

To answer RQ4, the paper provides a list of 20 features with the highest information gain along with their frequency in the projects. The vocabulary associated with flaky tests contains words such as \texttt{job} (present only in 2 projects (out of 24)), \texttt{id} (appears in 9 projects), \texttt{table}, and \texttt{action}, many of which are associated with executing tasks remotely and/or using an event queue.  Interestingly, the authors did not find a single token in the top 20 that was more strongly associated with non-flakiness. In that regard, for the Java keyword \texttt{throws} the authors conjecture that proper exception handling can help avoiding test flakiness.

\subsection{Threats and Limitations}

One of the main threats mentioned by the authors is the generalization of the results. The obtained vocabulary is limited to the Java language. But in addition to this, even if the same language is considered we would like to highlight two other threats. The first one is that the results may be valid only for the studied test cases, which are particular to the selected projects and their domains. A second possible threat is the ML algorithms used. 
To investigate the impact of such threats is the objective of our replication study. 

\section{Replication Study Setup}
\label{sec:setup}

We set out our replication study to address the threats of the original study mentioned in the last section. We can divide our study in two parts, each one of them corresponding to one of the identified threats and to a main research question. 

\vspace{0.4cm} \noindent \textbf{RQ$_1$}. Can we obtain similar results by using other ML algorithms or the same algorithms with distinct implementations?

\begin{itemize}
    \item \textbf{RQ$_{1.1}$} How accurately can we predict test flakiness based on source code identifiers in the test cases?
    
    \item \textbf{RQ$_{1.2}$} What value of different features add to the classifier?
    
    \item \textbf{RQ$_{1.3}$} Which test code identifiers are most strongly associated with test flakiness?
\end{itemize}

\textbf{Rationale.} 
Answering {RQ$_1$} we intend to investigate the influence of using other algorithms and implementations in the results of the original study.  We believe the previous results are promising, yet further evidence would be desirable.  For this end, replication studies are an important source of evidence~\cite{Campbell-Stanley:1963}. Then, apart from the first RQ of the original study that aims to obtain the dataset, we used the dataset \msrflakiness (available lab package) to replicate the study for the remaining questions. 
As such, RQ$_{1.1}$, RQ$_{1.2}$, and RQ$_{1.3}$ are same ones investigated previously, but now using the framework Scikit-learn and applying three additional algorithms, successfully employed for processing text.

\vspace{0.4cm} \noindent \textbf{RQ$_2$}. Are the results valid for other datasets including different projects?

\begin{itemize}
    \item \textbf{RQ$_{2.1}$} Can a trained classifier be successfully applied within the same projects (i.e., intra-project)?
    
    \item \textbf{RQ$_{2.2}$} Can a trained classifier be successfully applied to other projects (i.e., inter-projects)? 
\end{itemize}

\textbf{Rationale.} 
RQ$_2$ investigates if the vocabulary of flaky tests obtained by original study remains valid and can be directly used in other contexts.  Answering this question, we intend to obtain evidence that the results can be generalized by varying the datasets used. For this end, we use different cross-project validation, not only using the original \msrflakiness.  This question is herein proposed to assess the behavior of trained models towards the adoption by practitioners. In particular, we analyze two scenarios: \textit{(i)} (RQ$_{2.1}$) there is a group of projects and historical data is used to predict flaky tests within this group (i.e., intra-project), and \textit{(ii)} (RQ$_{2.2}$) to predict test flakiness in projects outside the group (i.e., inter-projects).

\vspace{0.2cm}
To answer \textbf{RQ$_1$}, we used the lab package \msrflakiness to replicate the original study in two steps. First like the work of Pinto et al., we adopted the ML framework Weka \cite{Witten99weka:practical} to extract features, training and validation of ML algorithms. This step tries to reproduce the same results, as well as to know the intrinsic configurations of feature extraction, training and tests. The idea is to confirm previous results and understand the available package \msrflakiness. 
As a result of this dependent replication, apart from RQ1, which is not directly related to the goal of  our study, we obtained the same results of the original study for all questions. These results are not presented and discussed in the present paper but they are in our replication package.

In this way, as a result of this first step  we  ensure that we  are interpreting well and adopting the same procedure of original study. Then in the second step and considering the goals of \textbf{RQ$_1$}, we use a different ML framework and algorithms, so we chose Scikit-learn~\cite{Pedregosa-etal:2011}. Scikit-learn is one of the most popular ML libraries\footnote{https://github.blog/2019-01-24-the-state-of-the-octoverse-machine-learning/}, providing a platform to build ML applications using the programming language Python. This makes it possible to evaluate if the results could be reproduced in a different software platform, and facilitates future extensions of the study, once Python seems to be the de facto programming language for ML~\cite{Goddard:2018}.

 Targeting RQ$_{1.1}$, we first loaded the dataset  \msrflakiness and ran the pre-processing of features (tokens). For feature extraction, we adopted the method \textsc{CountVectorizer} of Scikit-learn along with a handcrafted tokenizer, shown in \Cref{src:word-tokenizer}, to replicate the same tokenization criteria used in Weka at the original study. 

\begin{figure}[hb]
\begin{lstlisting}[language=Python, basicstyle=\scriptsize]
def weka_tokenizer(string):
    delimiters = 
        re.compile("[ |\n|\f|\r|\t|.|,|;|:|'|\"|(|)|?|!]")
    return list(filter(None, delimiters.split(string)))
\end{lstlisting}
\vspace*{-2mm}
\caption{Function for word separation in Python.}
\label{src:word-tokenizer}
\end{figure}

Scikit-learn has implementations of the five classifiers used in the original study: Random Forest, Decision Tree (DT), Naive Bayes, Support Vector Machine (SVM) and Nearest Neighbour.  Yet, we needed to adjust them to use the same parameters of Weka to obtain the same results. 
Furthermore, we included three different classifiers: Logistic Regression (LR),  Linear Discriminant Analysis (LDA) and Perceptron. The choice of these algorithms is based on successful studies that use linear classifiers and neural networks  for text classification~\cite{dreiseitl:2002, pranckevivcius:2017}.  

We used 80\% of the dataset for training and 20\% for validation, following the original study. \Cref{src:classifiers} shows how the classifiers were parameterized in Scikit-learn based on the original study Weka parameters. We used the \textsc{LinearSVC} implementation of SVM. 

\begin{figure}[hbt]
\begin{lstlisting}[language=Python, basicstyle=\scriptsize]
'random_forest': RandomForestClassifier(random_state=1), 
'decision_tree': DecisionTreeClassifier(min_samples_leaf=1),
'naiveb_bayes': GaussianNB(),
'svm': CalibratedClassifierCV(LinearSVC(fit_intercept=False, tol=0.001, C=1, dual=False, max_iter=100000), method='sigmoid'),
'knn': KNeighborsClassifier(n_neighbors=1, metric='euclidean'),
'LR': LogisticRegression(max_iter=1000),
'perceptron': CalibratedClassifierCV(Perceptron()),
'LDA': LinearDiscriminantAnalysis()
\end{lstlisting}
\vspace*{-2mm}
\caption{Classifiers' parameters used in Scikit-learn.}
\label{src:classifiers}
\end{figure}

Concerning RQ$_{1.2}$, the goal is to evaluate the impact of pre-processing steps for word separation and feature transformation. The following configurations were adopted considering datasets made available by the \msrflakiness experimental package: No Stemming, No Stop Words Removal, No Lowercasing, No Identifier Split, Only Split Identifiers, No Lines of Code, No Java Keywords, No Identifiers. These configurations were evaluated with the same algorithms of the original study to allow comparison.

In RQ$_{1.3}$, we evaluated the impact of features using the method  \texttt{mutual\_info\_classif} of Scikit-learn with default settings, which is equivalent to the entropy calculation of Weka. At this step we considered the entire dataset. The information gain (as known as entropy) is calculated for each output variable. This value ranges from 0 (no gain) to 1 (maximum of information gain).
All these aforementioned steps were performed as described in the previous work.

Concerning \textbf{RQ$_2$}, a different dataset is required to validate the trained models. Therefore, we selected the 335 flaky tests from 72 different projects, collected in \citet{Lam-etal:2019}; we refer to this dataset as iDFlakies. This dataset does not contain examples of non-flaky tests, so only recall will be used to analyze the  results. To obtain the dataset's features, we used the same process performed by the original study based on the scripts for data extraction. To support this step, we adopted the eight classifiers trained for answering RQ$_1$ using Scikit-learn. 

RQ$_{2.1}$ focuses on the intra-project scenario.  
For this question, we generated from iDFlakies a validation dataset using the process provided by the authors of the original study. As a result we transformed the flaky test cases into a dataset that comprises all the features used in the original study. Then we filtered out duplicate tests and tests from projects not present in dataset \msrflakiness. So, we are testing the scenario in which the historical data of a set of projects is used to predict test flakiness within this same set of projects. In the end, the validation set for this question contains only flaky tests (80) from 22 different projects. 

\begin{table*}[!b]
\small
\centering
\caption{Classifier performance.}
\label{tab:resultados}
\resizebox{\textwidth}{!}{%
\begin{tabular}{lllllllllllll}
\cline{1-6} \cline{8-13}
\multicolumn{6}{|c|}{\textbf{(a) Original study}} &
  \multicolumn{1}{l|}{} &
  \multicolumn{6}{c|}{\textbf{(b) Replication study}} \\ \cline{1-6} \cline{8-13} 
\multicolumn{1}{|l|}{\textbf{Algorithm}} &
  \multicolumn{1}{l|}{\textbf{Precision}} &
  \multicolumn{1}{l|}{\textbf{Recall}} &
  \multicolumn{1}{l|}{\textbf{F$_1$}} &
  \multicolumn{1}{l|}{\textbf{MCC}} &
  \multicolumn{1}{l|}{\textbf{AUC}} &
  \multicolumn{1}{l|}{} &
  \multicolumn{1}{l|}{\textbf{Algorithm}} &
  \multicolumn{1}{l|}{\textbf{Precision}} &
  \multicolumn{1}{l|}{\textbf{Recall}} &
  \multicolumn{1}{l|}{\textbf{F$_1$}} &
  \multicolumn{1}{l|}{\textbf{MCC}} &
  \multicolumn{1}{l|}{\textbf{AUC}} \\ \cline{1-6} \cline{8-13} 
\multicolumn{1}{|l|}{Random Forest} &
  \multicolumn{1}{l|}{\textbf{0.99}} &
  \multicolumn{1}{l|}{0.91} &
  \multicolumn{1}{l|}{\textbf{0.95}} &
  \multicolumn{1}{l|}{\textbf{0.90}} &
  \multicolumn{1}{l|}{\textbf{0.98}} &
  \multicolumn{1}{l|}{} &
  \multicolumn{1}{l|}{Random Forest} &
  \multicolumn{1}{l|}{\textbf{0.98}} &
  \multicolumn{1}{l|}{0.89} &
  \multicolumn{1}{l|}{\textbf{0.94}} &
  \multicolumn{1}{l|}{\textbf{0.89}} &
  \multicolumn{1}{l|}{\textbf{0.98}} \\ \cline{1-6} \cline{8-13} 
\multicolumn{1}{|l|}{Decision Tree} &
  \multicolumn{1}{l|}{0.89} &
  \multicolumn{1}{l|}{0.88} &
  \multicolumn{1}{l|}{0.89} &
  \multicolumn{1}{l|}{0.77} &
  \multicolumn{1}{l|}{0.91} &
  \multicolumn{1}{l|}{} &
  \multicolumn{1}{l|}{Decision Tree} &
  \multicolumn{1}{l|}{0.87} &
  \multicolumn{1}{l|}{0.86} &
  \multicolumn{1}{l|}{0.86} &
  \multicolumn{1}{l|}{0.74} &
  \multicolumn{1}{l|}{0.87} \\ \cline{1-6} \cline{8-13} 
\multicolumn{1}{|l|}{Naive Bayes} &
  \multicolumn{1}{l|}{0.93} &
  \multicolumn{1}{l|}{0.80} &
  \multicolumn{1}{l|}{0.86} &
  \multicolumn{1}{l|}{0.74} &
  \multicolumn{1}{l|}{0.93} &
  \multicolumn{1}{l|}{} &
  \multicolumn{1}{l|}{Naive Bayes} &
  \multicolumn{1}{l|}{0.95} &
  \multicolumn{1}{l|}{0.84} &
  \multicolumn{1}{l|}{0.89} &
  \multicolumn{1}{l|}{0.81} &
  \multicolumn{1}{l|}{0.90} \\ \cline{1-6} \cline{8-13} 
\multicolumn{1}{|l|}{SVM} &
  \multicolumn{1}{l|}{0.93} &
  \multicolumn{1}{l|}{\textbf{0.92}} &
  \multicolumn{1}{l|}{0.93} &
  \multicolumn{1}{l|}{0.85} &
  \multicolumn{1}{l|}{0.93} &
  \multicolumn{1}{l|}{} &
  \multicolumn{1}{l|}{SVM} &
  \multicolumn{1}{l|}{0.93} &
  \multicolumn{1}{l|}{0.86} &
  \multicolumn{1}{l|}{0.90} &
  \multicolumn{1}{l|}{0.81} &
  \multicolumn{1}{l|}{0.96} \\ \cline{1-6} \cline{8-13} 
\multicolumn{1}{|l|}{Nearest Neighbour} &
  \multicolumn{1}{l|}{0.97} &
  \multicolumn{1}{l|}{0.88} &
  \multicolumn{1}{l|}{0.92} &
  \multicolumn{1}{l|}{0.85} &
  \multicolumn{1}{l|}{0.93} &
  \multicolumn{1}{l|}{} &
  \multicolumn{1}{l|}{Nearest Neighbour} &
  \multicolumn{1}{l|}{\textbf{0.98}} &
  \multicolumn{1}{l|}{0.81} &
  \multicolumn{1}{l|}{0.89} &
  \multicolumn{1}{l|}{0.81} &
  \multicolumn{1}{l|}{0.90} \\ \cline{1-6} \cline{8-13} 
 &
   &
   &
   &
   &
   &
   &
   &
   &
   &
   &
   &
   \\ \cline{8-13} 
    &
   &
   &
   &
   &
   &
  \multicolumn{1}{l|}{} &
  \multicolumn{1}{l|}{Perceptron} &
  \multicolumn{1}{l|}{0.95} &
  \multicolumn{1}{l|}{0.83} &
  \multicolumn{1}{l|}{0.88} &
  \multicolumn{1}{l|}{0.81} &
  \multicolumn{1}{l|}{0.96} \\ \cline{8-13} 
 &
   &
   &
   &
   &
   &
  \multicolumn{1}{l|}{} &
  \multicolumn{1}{l|}{LR} &
  \multicolumn{1}{l|}{0.91} &
  \multicolumn{1}{l|}{\textbf{0.91}} &
  \multicolumn{1}{l|}{0.91} &
  \multicolumn{1}{l|}{0.84} &
  \multicolumn{1}{l|}{0.96} \\ \cline{8-13} 
 &
  &
  &
  &
  &
  &
  \multicolumn{1}{l|}{} &
  \multicolumn{1}{l|}{LDA} &
  \multicolumn{1}{l|}{0.83} &
  \multicolumn{1}{l|}{0.78} &
  \multicolumn{1}{l|}{0.80} &
  \multicolumn{1}{l|}{0.63} &
  \multicolumn{1}{l|}{0.87} \\ \cline{8-13} 
\end{tabular}%
}
\end{table*}

In RQ$_{2.2}$, we evaluated the inter-project scenario. 
So, we generated from iDFlakies a validation dataset by  filtering out tests from projects present in the dataset  \msrflakiness. 
So, we are testing the scenario in which the historical data of a set of projects is used to predict test flakiness in a different set of projects.
At the end, the validation set for this question contains only flaky tests (256) from 64 different projects. 
\section{Analysis of Results}
\label{sec:results}

This section analyses the obtained results in order to answer the posed RQs. 

\subsection{RQ$_1$ -- Replicating the study with Scikit-learn}

Below we compare the results of our replication using Scikit-learn with the ones obtained in the original study using Weka.  In our analysis we adopted the same evaluation metrics used in the original study (see Section~\ref{sec:original}): precision, recall, F$_1$-score, MCC and AUC. 

\subsubsection{RQ$_{1.1}$ -- How accurately can we predict test flakiness based on source code identifiers in the test cases?}

The results of our replication using Scikit-learn were similar to the results of the original study (which used Weka), as shown in \Cref{tab:resultados}. The greatest difference was for the Nearest Neighbour classifier, with a -0.7\% difference for Scikit-learn considering Recall, and Naive Bayes, with a 0.8\% difference for Scikit-learn regarding MCC. 

The Random Forest algorithm had the highest score for all evaluated metrics in our replication study considering the algorithms of the original study. It correctly identified 98\% of all the flaky tests classified by the model. The results for MCC of 0.90 and for F$_1$-score of 0.94 show that the quality of the prediction model is very good. The high score for both MCC (0.90) and F$_1$-score (0.94) were expected, as the dataset was balanced. AUC for Random Forest was also the highest (0.98), providing further evidence regarding the model classification quality. Considering the extended set of classifiers, Random Forest had a lower score only for Recall when considering the Logistic Regression classifier using the Scikit-learn: 0.90 versus 0.91, respectively.

For the new classifiers considered in our study, Logistic regression provided comparable results to Random Forest, with better results for recall. Nonetheless, the difference was of just two percentage points. Perceptron had a similar performance to Logistic Regression. Regarding LDA, it performed similarly to Decision Tree, with a lower overall performance than the other algorithms.

\begin{tcolorbox}
Answer to RQ$_{1.1}$: The classifiers performed very well, similarly to the original study. The difference of the results due to machine learning frameworks is small.  The additional classifiers obtained results similar to the ones investigated previously. LR presented the best recall value.
\end{tcolorbox}

\subsubsection{RQ$_{1.2}$  -- What value of different features add to the classifier?}

Although every classifier has performed very well, we considered models created with all the features extracted from the dataset. However, it is important to  analyse the impact of each feature and pre-processing applied to them. \citet{Pinto-etal:2020} considered three types of features: identifiers, Java keywords, and LOC metric. For the identifiers, which comprise most of the features considered in the model, several pre-processing steps were applied: stemming, stop word removal, lowercasing, splitting. In \Cref{tab:performance-without-features}, we present the performance of Random Forest and SVM classifiers with respect to precision, recall, F$_1$-score, MCC and AUC, with different subsets of features and pre-processing configurations.

As in the original study, the exclusion of tokens from the generated model (No Identifiers) had the greatest impact in the classifier performance in the replication. The precision reduces drastically, from 0.98 to 0.72 when using Random Forest and from 0.94 to 0.59 using SVM. Considering that identifiers are directly related to the vocabulary in each test case, this result was expected. Interestingly, disabling pre-processing for identifiers did not significantly changed the performance of the model. Similarly, excluding LOC or Java keywords  other features also have minor impact, with a variation of less than three percentage points for SVM. 

\begin{table*}[htb]
\small
\centering
\caption{Performance of RQ 1.1 classifiers without features.}
\label{tab:performance-without-features}
\resizebox{\textwidth}{!}{%
\begin{tabular}{|l|l|l|l|l|l|l|l|l|l|l|l|l|}
\cline{1-6} \cline{8-13}
\multicolumn{6}{|c|}{\textbf{Original study}}                    &  & \multicolumn{6}{c|}{\textbf{Replication study}}                  \\ \cline{1-6} \cline{8-13} 
\multicolumn{6}{|c|}{\textbf{(a) Random Forest}}                 &  & \multicolumn{6}{c|}{\textbf{(b) Random Forest}}                  \\ \cline{1-6} \cline{8-13} 
\textbf{Features} &
  \textbf{Precision} &
  \textbf{Recall} &
  \textbf{F$_1$} &
  \textbf{MCC} &
  \textbf{AUC} &
   &
  \textbf{Features} &
  \textbf{Precision} &
  \textbf{Recall} &
  \textbf{F1} &
  \textbf{MCC} &
  \textbf{AUC} \\ \cline{1-6} \cline{8-13} 
\textbf{All Features}         & 0.99 & 0.91 & 0.95 & 0.90 & 0.98 &  & \textbf{All Features}         & 0.98 & 0.90 & 0.94 & 0.90 & 0.98 \\ \cline{1-6} \cline{8-13} 
\textbf{No Stemming}          & 0.99 & 0.91 & 0.95 & 0.90 & 0.98 &  & \textbf{No Stemming}          & 0.97 & 0.89 & 0.93 & 0.88 & 0.98 \\ \cline{1-6} \cline{8-13} 
\textbf{No Stop W. Removal}   & 0.99 & 0.91 & 0.95 & 0.90 & 0.98 &  & \textbf{No Stop W. Removal}   & 0.98 & 0.89 & 0.93 & 0.87 & 0.98 \\ \cline{1-6} \cline{8-13} 
\textbf{No Lowercasing}       & 0.98 & 0.91 & 0.94 & 0.89 & 0.98 &  & \textbf{No Lowercasing}       & 0.98 & 0.90 & 0.93 & 0.88 & 0.98 \\ \cline{1-6} \cline{8-13} 
\textbf{No Identifier Split.} & 0.98 & 0.89 & 0.94 & 0.88 & 0.98 &  & \textbf{No Identifier Split.} & 0.98 & 0.90 & 0.94 & 0.89 & 0.98 \\ \cline{1-6} \cline{8-13} 
\textbf{Only Split Identif.}  & 0.99 & 0.92 & 0.95 & 0.90 & 0.98 &  & \textbf{Only Split Identif.}  & 0.97 & 0.90 & 0.93 & 0.88 & 0.98 \\ \cline{1-6} \cline{8-13} 
\textbf{No Lines of Code}     & 0.99 & 0.91 & 0.95 & 0.90 & 0.99 &  & \textbf{No Lines of Code}     & 0.98 & 0.89 & 0.93 & 0.88 & 0.98 \\ \cline{1-6} \cline{8-13} 
\textbf{No Java Keywords}     & 0.99 & 0.90 & 0.94 & 0.89 & 0.98 &  & \textbf{No Java Keywords}     & 0.98 & 0.89 & 0.93 & 0.88 & 0.98 \\ \cline{1-6} \cline{8-13} 
\textbf{No Identifiers}       & 0.76 & 0.82 & 0.79 & 0.56 & 0.85 &  & \textbf{No Identifiers}       & 0.72 & 0.81 & 0.76 & 0.52 & 0.85 \\ \cline{1-6} \cline{8-13} 
\multicolumn{6}{|c|}{\textbf{(c) SVM}}                           &  & \multicolumn{6}{c|}{\textbf{(d) SVM}}                            \\ \cline{1-6} \cline{8-13} 
\textbf{Features} &
  \textbf{Precision} &
  \textbf{Recall} &
  \textbf{F1} &
  \textbf{MCC} &
  \textbf{AUC} &
  \textbf{} &
  \textbf{Features} &
  \textbf{Precision} &
  \textbf{Recall} &
  \textbf{F1} &
  \textbf{MCC} &
  \textbf{AUC} \\ \cline{1-6} \cline{8-13} 
\textbf{All Features}         & 0.93 & 0.92 & 0.93 & 0.85 & 0.93 &  & \textbf{All Features}         & 0.94 & 0.86 & 0.90 & 0.82 & 0.96 \\ \cline{1-6} \cline{8-13} 
\textbf{No Stemming}          & 0.93 & 0.92 & 0.93 & 0.85 & 0.93 &  & \textbf{No Stemming}          & 0.94 & 0.85 & 0.89 & 0.81 & 0.96 \\ \cline{1-6} \cline{8-13} 
\textbf{No Stop W. Removal}   & 0.93 & 0.92 & 0.93 & 0.85 & 0.93 &  & \textbf{No Stop W. Removal}   & 0.93 & 0.86 & 0.89 & 0.81 & 0.96 \\ \cline{1-6} \cline{8-13} 
\textbf{No Lowercasing}       & 0.91 & 0.93 & 0.92 & 0.84 & 0.92 &  & \textbf{No Lowercasing}       & 0.93 & 0.86 & 0.90 & 0.81 & 0.96 \\ \cline{1-6} \cline{8-13} 
\textbf{No Identifier Split.} & 0.91 & 0.88 & 0.89 & 0.79 & 0.90 &  & \textbf{No Identifier Split.} & 0.94 & 0.88 & 0.91 & 0.84 & 0.97 \\ \cline{1-6} \cline{8-13} 
\textbf{Only Split Identif.}  & 0.93 & 0.92 & 0.93 & 0.85 & 0.93 &  & \textbf{Only Split Identif.}  & 0.92 & 0.90 & 0.91 & 0.83 & 0.96 \\ \cline{1-6} \cline{8-13} 
\textbf{No Lines of Code}     & 0.93 & 0.92 & 0.93 & 0.85 & 0.93 &  & \textbf{No Lines of Code}     & 0.94 & 0.86 & 0.90 & 0.80 & 0.96 \\ \cline{1-6} \cline{8-13} 
\textbf{No Java Keywords}     & 0.93 & 0.92 & 0.93 & 0.85 & 0.93 &  & \textbf{No Java Keywords}     & 0.92 & 0.89 & 0.91 & 0.82 & 0.96 \\ \cline{1-6} \cline{8-13} 
\textbf{No Identifiers}       & 0.54 & 0.87 & 0.74 & 0.40 & 0.68 &  & \textbf{No Identifiers}       & 0.59 & 0.83 & 0.69 & 0.32 & 0.73 \\ \cline{1-6} \cline{8-13} 
\end{tabular}%
}
\end{table*}

\vspace{.3cm}
\begin{tcolorbox}
Answer to RQ$_{1.2}$: Similarly to the original study, different features did not show much impact in the classifiers, except when tokens (No Identifiers)  extracted from test cases were not considered. 
\end{tcolorbox}

\subsubsection{RQ$_{1.3}$  -- Which test code identifiers are most strongly associated with test flakiness?}

The association of features and test code identifiers was evaluated by calculating the information gain based on the entropy of the features extracted after pre-processing the text of the test cases. In \Cref{tab:top-features-inf-gain} we report the features with higher information gain for original study and our replication. The value for the information gain of the studies are different. For instance, for \texttt{job} the original study reported 0.2053 while we got 0.1449 for the replication. Nonetheless, the top three features are the same for both models and, considering the top twenty features, the only differences are the feature \texttt{coordinatorjob} (present in the original study, but absent from the replication) and \texttt{get} (present in the replication, but absent in the original study).

Thus, despite the different values for information gain, we can conclude that features associated with test flakiness are the mostly same of the original study. The features are associated with executing and coordinating tasks (\texttt{job}, \texttt{services}, \texttt{action}, \texttt{coordinator}, \texttt{workflow}, \texttt{getstatus}), persistance (\texttt{table}, \texttt{id}, \texttt{record}, \texttt{jpa}, \texttt{jpaservice}). The most relevant difference was with respect to the feature \texttt{throws (keyword)}. Although it was identically ranked, in the original study it was related to few tests (10), in the replication it was exactly the opposite, with 2.348 test cases. We also conjecture that exception handling is related to flakiness.

\begin{table*}[htb]
\small
\centering
\caption{Top 20 features by Information Gain.}
\label{tab:top-features-inf-gain}
\resizebox{\textwidth}{!}{%
\begin{tabular}{|l|l|l|l|l|l|l|l|l|l|l|}
\cline{1-5} \cline{7-11}
\multicolumn{5}{|c|}{\textbf{(a) Original study}} &
   &
  \multicolumn{5}{c|}{\textbf{(b) Replication study}} \\ \cline{1-5} \cline{7-11} 
\textbf{Feature} &
  \textbf{Inf. Gain} &
  \textbf{Tests} &
  \textbf{\begin{tabular}[c]{@{}l@{}}Flaky\\ tests\end{tabular}} &
  \textbf{\begin{tabular}[c]{@{}l@{}}Non-Flaky\\ tests\end{tabular}} &
   &
  \textbf{Feature} &
  \textbf{Inf. Gain} &
  \textbf{Tests} &
  \textbf{\begin{tabular}[c]{@{}l@{}}Flaky\\ tests\end{tabular}} &
  \textbf{\begin{tabular}[c]{@{}l@{}}Non-Flaky\\ tests\end{tabular}} \\ \cline{1-5} \cline{7-11} 
job &
  0.2053 &
  528 &
  524 &
  4 &
   &
  job &
  0.1449 &
  530 &
  525 &
  5 \\ \cline{1-5} \cline{7-11} 
table &
  0.1449 &
  414 &
  406 &
  8 &
   &
  table &
  0.1029 &
  414 &
  406 &
  8 \\ \cline{1-5} \cline{7-11} 
id &
  0.1419 &
  584 &
  522 &
  62 &
   &
  id &
  0.1004 &
  577 &
  525 &
  52 \\ \cline{1-5} \cline{7-11} 
action &
  0.1365 &
  395 &
  387 &
  8 &
   &
  services &
  0.0977 &
  378 &
  371 &
  7 \\ \cline{1-5} \cline{7-11} 
oozie &
  0.1359 &
  274 &
  274 &
  0 &
   &
  action &
  0.0972 &
  396 &
  388 &
  8 \\ \cline{1-5} \cline{7-11} 
services &
  0.1309 &
  378 &
  371 &
  7 &
   &
  oozie &
  0.0942 &
  346 &
  346 &
  0 \\ \cline{1-5} \cline{7-11} 
coord &
  0.1192 &
  307 &
  307 &
  0 &
   &
  loc &
  0.0879 &
  \multicolumn{1}{c|}{-} &
  \multicolumn{1}{c|}{-} &
  \multicolumn{1}{c|}{-} \\ \cline{1-5} \cline{7-11} 
getid &
  0.1076 &
  288 &
  287 &
  1 &
   &
  coord &
  0.0826 &
  307 &
  307 &
  0 \\ \cline{1-5} \cline{7-11} 
coordinator &
  0.1070 &
  258 &
  258 &
  0 &
   &
  xml &
  0.0752 &
  356 &
  341 &
  15 \\ \cline{1-5} \cline{7-11} 
xml &
  0.1061 &
  253 &
  247 &
  6 &
   &
  getid &
  0.0746 &
  288 &
  287 &
  1 \\ \cline{1-5} \cline{7-11} 
loc &
  0.0977 &
  \multicolumn{1}{c|}{-} &
  \multicolumn{1}{c|}{-} &
  \multicolumn{1}{c|}{-} &
   &
  coordinator &
  0.0741 &
  278 &
  278 &
  0 \\ \cline{1-5} \cline{7-11} 
workflow &
  0.0913 &
  207 &
  207 &
  0 &
   &
  get &
  0.0691 &
  2194 &
  1260 &
  934 \\ \cline{1-5} \cline{7-11} 
getstatus &
  0.0884 &
  248 &
  246 &
  2 &
   &
  workflow &
  0.0633 &
  240 &
  240 &
  0 \\ \cline{1-5} \cline{7-11} 
throws\_keyword &
  0.0873 &
  10 &
  3 &
  7 &
   &
  throws\_keyword &
  0.0615 &
  2348 &
  1327 &
  1021 \\ \cline{1-5} \cline{7-11} 
record &
  0.0845 &
  314 &
  296 &
  18 &
   &
  getstatus &
  0.0613 &
  248 &
  246 &
  2 \\ \cline{1-5} \cline{7-11} 
jpa &
  0.0780 &
  207 &
  207 &
  0 &
   &
  record &
  0.0596 &
  314 &
  296 &
  18 \\ \cline{1-5} \cline{7-11} 
jpaservice &
  0.0752 &
  200 &
  200 &
  0 &
   &
  service &
  0.0590 &
  451 &
  383 &
  68 \\ \cline{1-5} \cline{7-11} 
service &
  0.0733 &
  434 &
  367 &
  67 &
   &
  jpa &
  0.0541 &
  207 &
  207 &
  0 \\ \cline{1-5} \cline{7-11} 
wf &
  0.0721 &
  192 &
  192 &
  0 &
   &
  jpaservice &
  0.0521 &
  200 &
  200 &
  0 \\ \cline{1-5} \cline{7-11} 
coordinatorjob &
  0.0689 &
  184 &
  184 &
  0 &
   &
  wf &
  0.0499 &
  192 &
  192 &
  0 \\ \cline{1-5} \cline{7-11} 
\end{tabular}%
}
\end{table*}

\vspace{.3cm}
\begin{tcolorbox}
Answer to RQ$_{1.3}$: The flaky vocabulary is related to executing and coordinating tasks, and persistence. The set of words is very similar to the one of original study.
\end{tcolorbox}

\vspace{0.3cm}
\noindent {\bf {\it Summary of RQ$_{1}$}}: Our replication gives more evidence about the validity of the original results using other ML framework and algorithms. The results obtained are very similar, concerning the performance of the algorithms, value of the used features and vocabulary of flaky tests found. It is worth emphasizing  the good performance of the additional classifiers. LR reached the best values of Recall and second best values of AUC. Other point to be highlighted is that the lab package provided by the original study is easy-to-use and self-contained, which allows us to  replicate the study without contacting the authors.

\subsection{RQ$_2$ --  Cross-project validation}

For RQ$_2$, we extended the original study by addressing the performance of the classification model using the vocabulary of flaky tests with different datasets. First, we consider using the trained classifier within a different set of test cases from the same software projects (intra-projects). Second, we applied the classifier to other projects (inter-projects).

\subsubsection{RQ$_{2.1}$ --  Can a trained classifier be successfully applied within the same projects (i.e., intra-project)?}

The intra-project validation was performed with the dataset of iDFlakies considering only the 22 projects which are  also in the dataset \msrflakiness. The results are presented in \Cref{tab:validation-intra-projects}. As previously described, this dataset does not have tests classified as non-flaky, thus the confusion matrix has only true positives (TP) and false negatives (FN). Thus, we just considered Recall to evaluate the performance of the classifiers.

The classifiers performance was not satisfactory compared to the results for \msrflakiness in the RQ$_{1.1}$ (\Cref{tab:resultados}). Random Forest was the classifier with the best performance regarding Recall on \msrflakiness (0.90), but achieved only 0.08 for the intra-project scenario. SVM also experienced a lower recall, dropping from 0.86 to 0.29. Interestingly, the classifiers we had added for the extended replication performed better than the original ones, although they also had a lower performance. Linear Discriminant Analysis (LDA) had a recall of 0.75, followed by Logistic Regression (LR), with 0.68.

\begin{table}[htb!]
\small
\centering
\caption{Classifiers performance in the intra-project scenario.}
\label{tab:validation-intra-projects}
\begin{tabular}{|l|l|l|l|l|l|l|l|}
\hline
\textbf{Classifier} & \textbf{Recall} & \textbf{TP} & \textbf{FN} \\ \hline
LDA           & 0.75 & 60 & 20 \\ \hline
LR            & 0.68 & 54 & 26 \\ \hline
Perceptron    & 0.51 & 41 & 39 \\ \hline
SVM           & 0.29 & 23 & 57 \\ \hline
Decision Tree & 0.19 & 15 & 65 \\ \hline
KNN           & 0.19 & 15 & 65 \\ \hline
Random Forest & 0.08 & 6  & 74 \\ \hline
Naive Bayes   & 0.11 & 9  & 71 \\ \hline
\end{tabular}
\end{table}

Regarding the importance of each feature of the classification, we calculated the information gain and reported the top-20 features in the \Cref{tab:rq21-top-features-inf-gain}. The only feature common to \msrflakiness and the intra-project dataset was \texttt{loc}, but with no actual information. There some features related to execution and coordination of tasks (\texttt{init}, \texttt{createdirwithhttp}), but most are related to I/O operations (\texttt{reader}, \texttt{write}, \texttt{directory}, \texttt{folder}). The feature \texttt{public}, a Java keyword, had the highest information gain (0.8188) and is associated to 80 flaky tests. As most (if not all) tests are declared in public methods, that is not actually unexpected, although it should not be relevant to detect flaky tests. Another interesting fact is that, although we reported top-20 features in \Cref{tab:rq21-top-features-inf-gain}, only 15 had an information gain greater than zero: the five features with no information gain were LOC and Java keywords (\texttt{abstract}, \texttt{assert}, \texttt{boolean} and \texttt{break}). 

\begin{table}[htb!]
\small
\centering
\caption{Top 20 features by information gain in the intra-project scenario.}
\label{tab:rq21-top-features-inf-gain}
\begin{tabular}{|l|l|l|}
\hline
\textbf{Feature}   & \textbf{Inf. Gain} & \textbf{\begin{tabular}[c]{@{}l@{}}Tests\\ Flaky\end{tabular}} \\ \hline
public\_keyword    & 0.8188 & 80 \\ \hline
acl                & 0.8188 & 6  \\ \hline
created            & 0.8188 & 6  \\ \hline
reader             & 0.8188 & 6  \\ \hline
createdirwithhttp  & 0.0791 & 6  \\ \hline
createsnapshot     & 0.0791 & 6  \\ \hline
directory          & 0.0791 & 6  \\ \hline
folder             & 0.0791 & 6  \\ \hline
gethadoopusers     & 0.0791 & 6  \\ \hline
hadoopusersconftesthelper & 0.0791 & 6 \\ \hline
init               & 0.0791 & 6  \\ \hline
no                 & 0.0791 & 6  \\ \hline
touri              & 0.0791 & 6  \\ \hline
write              & 0.0791 & 6  \\ \hline
assertnotnull      & 1.1102 & 28 \\ \hline
loc                & 0.0    & -  \\ \hline
abstract\_keyword  & 0.0    & 0  \\ \hline
assert\_keyword    & 0.0    & 68 \\ \hline
boolean\_keyword   & 0.0    & 0  \\ \hline
break\_keyword     & 0.0    & 0  \\ \hline
\end{tabular}%
\end{table}

\vspace{.3cm}
\begin{tcolorbox}
Answer to RQ$_{2.1}$: The performance of the classification model to identify flaky tests within the same projects was low. The features with higher information gain are distinct from those previously identified, and they are related to I/O operations, execution and coordination of tasks, and Java keywords. 
\end{tcolorbox}

\subsubsection{RQ$_{2.2}$ -- Can a trained classifier be successfully applied to other projects (i.e., inter-projects)?}

To answer RQ$_{2.2}$, we used the classification models trained for RQ$_{1.1}$ to test the inter-project dataset. This dataset has 256 flaky tests of 64 projects distinct from those from \msrflakiness. Similarly to RQ$_{2.1}$, we considered only recall to evaluate the classifier performance as the inter-project dataset contains just flaky tests. 

The performance of the classifiers for the inter-projects dataset was very low, as detailed in  \cref{tab:testRQ22}. Considering the Decision Tree classifier, which achieved a recall of 0.86 in $RQ_{1.1}$, the result for the inter-project dataset was of 0.16. Yet, from the five models considered in the original study, it was the best classifier. Random Forest, which had a recall of 0.90 for in $RQ_{1.1}$, got only 0.02, correctly identifying just four flaky tests. In the context of our extended replication study, LDA was the best, with a recall of 0.48.

It is worth noticing that the performance for inter-projects was significantly lower than for the intra-project scenario. For instance, LDA had a recall of 0.76 in the latter against 0.48 in the former. 

\begin{table}[h]
\small
\centering
\caption{Classifiers performance in the inter-project scenario.}
\label{tab:testRQ22}
\begin{tabular}{|l|l|l|l|l|l|l|l|}
\hline
\textbf{Classifier} & \textbf{Recall} & \textbf{TP} & \textbf{FN} \\ \hline
LDA             & 0.48 & 122 & 134 \\ \hline
LR              & 0.40 & 102 & 154 \\ \hline
Perceptron      & 0.38 & 97  & 159 \\ \hline
Decision Tree   & 0.16 & 40  & 216 \\ \hline
SVM             & 0.11 & 27  & 229 \\ \hline
Naive Bayes     & 0.09 & 24  & 232 \\ \hline
KNN             & 0.09 & 22  & 234 \\ \hline
Random Forest   & 0.02 & 4   & 252 \\ \hline
\end{tabular}
\end{table}

The information gain for the features extracted from the inter-project dataset is shown in \Cref{tab:rq22-top-features-inf-gain}. The features are different from those of $RQ_{1.1}$ and intra-project dataset in $RQ_{2.1}$. The features that are more often associated to flaky tests are related to asynchronous calls (\texttt{await}, \texttt{export}, \texttt{handler}, \texttt{protocol}, \texttt{server}, \texttt{task}, \texttt{taskpayloadbuilder}, \texttt{url}) or they are Java keywords (\texttt{return}, \texttt{try}). Strikingly, the special character \texttt{\{} is among the top-20 features, even though it is present in every flaky test and could not provide any relevant information to classify a test case as flaky.

\begin{table}[htb!]
\small
\centering
\caption{Top 20 features by Information Gain in the inter-project scenario.}
\label{tab:rq22-top-features-inf-gain}
\begin{tabular}{|l|l|l|}
\hline
\textbf{Feature} & \textbf{Inf. Gain} & \textbf{\begin{tabular}[c]{@{}l@{}}Tests\\ Flaky\end{tabular}} \\ \hline
getname            & 0.8882 & 22 \\ \hline
namingcontext      & 0.8882 & 22 \\ \hline
same               & 0.8882 & 22 \\ \hline
await              & 0.8674 & 14 \\ \hline
export             & 0.8674 & 14 \\ \hline
return\_keyword    & 0.8465 & 17 \\ \hline
be                 & 0.8465 & 17 \\ \hline
handler            & 0.8465 & 30 \\ \hline
is                 & 0.8465 & 17 \\ \hline
this\_keyword      & 0.8396 & 14 \\ \hline
protocol           & 0.8396 & 14 \\ \hline
\}                 & 0.8396 & 14 \\ \hline
collections        & 0.8188 & 17 \\ \hline
context            & 0.8188 & 19 \\ \hline
server             & 0.8119 & 14 \\ \hline
task               & 0.8049 & 19 \\ \hline
of                 & 0.7633 & 19 \\ \hline
taskpayloadbuilder & 0.7633 & 19 \\ \hline
url                & 0.6939 & 40 \\ \hline
try\_keyword       & 0.6661 & 51 \\ \hline
\end{tabular}%
\end{table}

\vspace{.3cm}
\begin{tcolorbox}
Answer to RQ$_{2.2}$: Classifiers trained with the dataset \msrflakiness could not provide a satisfactory performance when testing a dataset of different projects (inter-projects). The features with higher information gain are related to asynchronous calls and they are distinct from those previously identified. 
\end{tcolorbox}

\noindent {\bf {\it Summary of RQ$_{2}$}}: We obtained some evidence that the vocabulary of flaky tests from original study cannot be directly used in other contexts. In general, training the classifiers with the dataset  \msrflakiness does not lead  a good prediction in both scenarios evaluated:  intra- and inter-projects. A possible reason for this is  overfitting, caused by excessive number of tokens in the vocabulary and few examples.

\section{Threats to Validity}
\label{sec:threats}

Threats to construct validity are related to metrics used to evaluate the results. As a replication, the same metrics of \citet{Pinto-etal:2020} were employed to support the comparisons. 
As such, this study is subjected to the same threats that are the evaluation method based on precision and F$_1$-Score. For RQ$_{2}$, a dataset only with flaky tests was used. So, the precision was always the maximum and we did not consider it in our analyses. This limitation should be addressed in future work.

Threats to internal validity may comprise the results when relating independent and dependent variables. 
To mitigate this, we carefully replicate the experiments so that the same results were achieved 
with the Weka software. 
With Scikit-learn, we tried our best to replicate the results, while slightly differences were noticed due to mostly the feature extraction. 

External validity is connected to the generalization of obtained results. 
While the replication brings more evidence and the adoption of different datasets helps to evaluate different scenarios, as in the original study we cannot generalize the results, we are limited to a Java language and limited domain projects. 
When looking at the intra-project and inter-project contexts, the results herein presented had a meaningful reduction of performance, the fact that motivates further research. The sample used to evaluate the classifiers is small and should be increased to better understand the possibilities.

\section{Discussion}
\label{sec:discussion}

This section discusses some implications of our findings, which may guide new research in the area.

The use of tokens extracted from test cases was effective in identifying flaky tests for the first part of our study. However, for the extended replication, considering test cases for the same projects (which should share the vocabulary) and for different projects (which may not hold such assumption), the recall was rather low. This suggests an overfitting of the classification model. 

The choice of features or models that are more generalizable and useful in broader contexts could be an alternative to improve the performance of classification models for flaky tests on intra and inter-projects scenarios. Several approaches could be analyzed to address this issue. For instance, we could refine a more general vocabulary, removing project-specific words. 

As reported, stop words had a minor impact on the classifier's performance. However, that could be due to an ineffective list of words. As we can see in  \Cref{tab:rq22-top-features-inf-gain}, there are features which are usually considered as stop words, such as common auxiliary verbs (\texttt{be}, \texttt{is}), prepositions (\texttt{of}), and symbols (\texttt{\}}). We could also employ term weighting (for example, TF-IDF) to reduce the impact of features common to a single project or to the entire dataset.

After inspecting the extracted features and the related test cases, we observed that the current tokenizer is splitting dates, URLs, and filenames. For usual text mining and NLP-based approaches, such heuristic would have minor impact on the classifier performance, but that information is relevant for flaky tests, which can be associated to I/O and time/calendar operations~\cite{Luo-etal:2014, WingLam-etal:2019:issta}. An improved parser and tokenizer could address such cases. 
Some tokens, like symbol ``\{'' in \Cref{tab:rq22-top-features-inf-gain}, being recognized as a feature point to issues regarding feature engineering.
It is important to emphasize that such special cases are also present in the original study.

Currently, we extracted features only from the body of the test cases, but tests can become flaky due to changes unrelated to that information~\cite{WingLam-etal:2020}. Thus we could consider external information associated to each test case, such as helper methods and library dependencies. Order-dependent flaky tests are often detected in test suites with helper methods~\cite{Shi-etal:2019} that configure the state of the application. For Java applications these fixture methods are annotated with \texttt{BeforeEach}, \texttt{BeforeClass}, \texttt{AfterEach}, and \texttt{AfterClass}, defining the test workflow. The association of such annotations to order-dependent flaky tests makes annotations a candidate for feature selection. Approximately 50\% of the flaky tests for the iDFlakies dataset considered for RQ$_{2.2}$ are order-dependent~\cite{Lam-etal:2019}. The absence of features associated to test workflow can partially explain the poor performance of the classifier for that dataset. 

Version control systems may also be an interesting source of data, such as code churn, modified code, commits, and contributors. 
Finally, dynamic features like code coverage, monitoring data, and test execution reports deserve further investigation.

\section{Conclusions}
\label{sec:conclusions}

Flaky tests are intermittently passing or failing, causing distrust in test automation. This may harm the software development process of several companies that rely on automated tests to support a continuous integration and delivery environment.
Therefore, the prevention and identification of potential flaky tests are investigated by practitioners and researchers. 

This paper intends to increase the body of evidence on using code identifiers to predict flaky tests. To do so, we conducted an extended replication of \citet{Pinto-etal:2020}' study on the vocabulary of flaky tests. 
Besides replicating the previous results, we extended it by using a different ML framework (namely, Scikit-learn), assessing different classifiers, and validating the trained models with different datasets for intra- and inter-project contexts. 

The results obtained in this work demonstrated that the proposal to create a vocabulary of flaky tests by the authors of the original study has a process of extracting characteristics, training is replicable and extensible, thus reducing possible bias added by researchers, process and software. The results obtained during the intra- and inter-project tests demonstrated that the defined vocabulary does not have the level of generalization sufficient to predict flaky tests with the same performance obtained during the tests of the classifiers. 
We observed that the context of a given project has a major impact on the vocabulary of flaky tests; this may hinder the adoption of code identifiers to predict test flakiness. 

A potential future work would be to investigate a subset of existing vocabularies that generalizes for different projects.  
Other direction is to evaluate if the prediction of test flakiness may benefit from different features used in other studies~\cite{Atif-etal:2017, TariqKing-etal:2018}, along with code identifiers. 

\section*{Acknowledgment}
This work is partially supported by CNPq (Andre T. Endo grant nr. 420363/2018-1 and Silvia Regina Vergilio grant nr. 305968/2018-1) and by IN2 Institute (Bruno Henrique Pachulski Camara grant nr. 002/2021).

\balance
\bibliographystyle{IEEEtranN}
\bibliography{ms}

\end{document}